\documentclass[conference]{IEEEtran}
\IEEEoverridecommandlockouts
\usepackage{cite}
\usepackage{amsmath,amssymb,amsfonts}
\usepackage{algorithmic}
\usepackage{graphicx}
\usepackage{textcomp}
\usepackage{xcolor}
\def\BibTeX{{\rm B\kern-.05em{\sc i\kern-.025em b}\kern-.08em
    T\kern-.1667em\lower.7ex\hbox{E}\kern-.125emX}}
\begin{document}

\title{Detecting Malicious Domains Using Statistical Internationalized Domain Name Features in Top Level Domains\\
}

\author{\IEEEauthorblockN{Alshaima Almarzooqi\IEEEauthorrefmark{1}, Jawahir Mahmoud\IEEEauthorrefmark{1}, Bayena Alzaabi \IEEEauthorrefmark{1}, Arsiema Ghebremichael \IEEEauthorrefmark{1} and Monther Aldwairi\IEEEauthorrefmark{1} }
\IEEEauthorblockA{\textit{\IEEEauthorrefmark{1}College of Technological Innovation} \\
\textit{Zayed University}\\
Abu Dhabi, UAE \\
{\{201507555, 201415600, 201500404, M80008144, monther.aldwairi\}@zu.ac.ae }}
}

\maketitle

\begin{abstract}
The Domain Name System (DNS) is a core Internet service that translates domain names into IP addresses. It is a distributed database and protocol with many known weaknesses that subject to countless attacks including spoofing attacks, botnets, and domain name registrations. Still, the debate between security and privacy is continuing, that is DNS over TLS or HTTP, and the lack of adoption of DNS security extensions, put users at risk. Consequently, the security of domain names and characterizing malicious websites is becoming a priority. This paper analyzes the difference between the malicious and the normal domain names and uses Python to extract various malicious DNS identifying characteristics. In addition, the paper contributes two categories of features that suppers Internationalized Domain Names and scans domain system using five tools to give it a rating. The overall accuracy of the Random Forest Classifier was 95.6\%.

\end{abstract}

\begin{IEEEkeywords}
DNS, malicious domains, malware detection, Internationalized Domain Name, URL, machine learning
\end{IEEEkeywords}

\section{Introduction}
The dramatic increase of Internet usage in the last two decades resulted in huge increase in domain names' registrations. Due to the very large number of domain names, it is very difficult to be able to distinguish and identify malicious ones from these that are authentic and safe. In addition, the introduction of the Internationalized Domain Names in Applications (IDNA) \cite{Klensin2010DN} to handle non-English domain names such as Arabic, Devanagari, Cyrillic, Chinese, and other languages to be encoded in Unicode. This made it harder to recognize and validate domain names. Therefore, it has become essential for domain registrars and users browsing the Internet, alike,to recognize malicious domains. There has been a lot of research on pinpointing malicious domains using URL features and artificial intelligence without the need to access, download and excuse the scripts hidden in the malicious website \cite{pope2012domain}.

We previously proposed lightweight machine learning-based methods for detection of Malicious URLs (MALURLS) \cite{Aldwairi2012MALURLSAL} \cite{Aldwairi2011MALURLsMU}. The previous work proofed the efficacy of using only URL features in malicious domain detection. More recent work demonstrated a cyber-threat alarm system for detecting malicious domains \cite{10.1007/978-3-319-67837-5_17}, detection of drive-by-downloads and malicious domains \cite{978-981-15-4825-3_6}, and detection of malicious advertisements using URL features \cite{7921994}. 

This paper presents an intelligent method for the identification and detection of DNS based malware using statistical and characterization features in both DNS and IDN domains. The proposed method will be able to use multiple behavioral features in order to discern malicious domains from the benign ones without accessing the domain or browsing the website. This is achieved by extending previous work  in \cite{messabi2018malware} by adding new features that covers IDN as well as classical DNS domains. The objective of the malware detection method is to detect and reduce malicious web attacks, by identifying potential threats early based on special DNS and IDN features of domains with high risks. 

Before we embark on the challenge detection and classification, we collect malicious domain names from various online available databases, such as DNS-BH Malware Domain Blocklist by RiskAnalytics  \cite{RiskAnalytics2019}, and PhishTank \cite{PhishTank2019}. Two whitelisted and blacklisted domains lists are selected and studied. Carefully selected features are then extracted using Python scripts.
Several tools are invoked to gather the calculate the feature values such as \textit{nslookup}, which was used to resolve the domains IP addresses. A list of characteristics and statistical features is used to represent or tag both valid and potentially risky domains. Then, once a domain is to be checked, the same features are extracted and evaluated for that particular domain, and hence, the domain can be classified as either safe or a potential threat. Finally, we experiment with several classifiers and choose the best performing classifier (Random Forest). 

The rest of this paper is organized as follows. Section II briefly surveys the literature on work that detects malicious domains. Section III presents the proposed DNS classification and identification method based on predefined features including new IDA features. Section IV presents the experimental evaluation of the proposed technique. Finally, conclusions and future work are discussed in Section V. 

\section{Related Work}
Domain name resolution is the first step before visiting any website and attackers historically, capitalized on the plain text and trust-based nature of the name resolution protocol and infrastructure. Undoubtedly, there has been an increasing research interest in detecting malicious pages before they are downloaded and browsed. Despite the numerous attempts, they e failed to provide a perfect solution with no deficiencies \cite{agyepong2018detection}. Snort is a popular signature-based intrusion detection system (IDS) that relies on pre-drafted rules to detect dubious network traffic and malware activity. The system focuses on detection of malware infections, that is the majority of the rules are essentially malware rules, and some are blacklist rules \cite{7209081,10.1002ett.3711}. More recently, Vulnerability Research Teams (VRT) started using machine learning to perform anomaly detection and rule generation for malicious DNS domains \cite{Aldwairi2018IRECAP}. However, the abundance of rule's sources and the fact that many are frequently manually updated, this made many rulesets unusable. Unfortunately, several rules had to be disabled or changed by hand before they will be deployed, and these modifications need to be re-done each time a new ruleset becomes available \cite{zhao2015detecting}.

According to the mining DNS for malicious domain registration research team, it seems that a good number of the newly registered domain names are malicious. They observed that benign domains names included proper English words and look a likes, while most of the malicious domains contain randomly generated and non meaningful words. They used second order Markov models along with a dictionary for the common English words, and transition matrices composed of known legitimate and malicious domain names. They used a Random Forest classifier with features extracted from DNS data, to detect malicious domains with low false positives \cite{he2010mining}.

Kopis monitored streams of DNS queries and responses from the higher DNS hierarchy, and detected malware domain names based on the query/response patterns. Kopis had 2 phases: a training mode and an operation mode. In training mode, Kopis made use of a domain knowledge base (KB), which consists of a group of far-famed malware-related and known legitimate domain names (and connected resolved IPs) Then in the operation mode it monitored the Authoritative Name Servers (AuthNS) and Top Level Domain (TLD) servers are authoritativeness and degree of delegation to detect malicious domains. However, it solely collected the traffic at the top-level-domains, therefore an access to authority-level DNS traffic needs to be granted \cite{antonakakis2011detecting}.

The "EXPOSURE" tool research team analyzed 100 billion DNS queries of 4.8 million domain names to prepare a blacklist. They observed and analyzed Alexa's most popular global sites queries to find the malicious domain activities. Their assumptions were that malicious services reply on the DNS structure, and that benign and malicious DNS requests do exhibit different behavior. They used 15 features extracted from DNS traffic to characterize malicious domains. to reduce the amount of data, they first filtered out the popular websites to reduce the amount of queries to examine to 20\%. Secondly, they filtered out domain names that are older than one year, on the premise that malicious domain names appear and shutdown after short period of time, therefore reduce queries by 50\%. During the 2.5 month analysis they managed to identify 3000 new malicious domains that were previously unknown \cite{bilge2011exposure}.

Aldwairi and Alsalman proposed MALURLs, a light weight statistical selflearning malicious domains detection technique based on URL lexical, host and special features. Host features included Internet protocol address, location, TTL and DNS records, while lexical featured included TLD length, URL length, number of dots and more. On the other hand special features where introduced to increase the accuracy such as JavaScript enable/disable, title and term frequency. The used 100 benign websites from Alexa and100 malicious sites from PhishTank. Due to the small size of the test dataset they used Genetic Algorithm to expand the dataset and performed classification using Naïve Bayes. They reported an average precision of  of 87\% \cite{Aldwairi2012MALURLSAL} \cite{Aldwairi2011MALURLsMU}.

Ghafir and Prenosil analyzed DNS traffic based in an automatically updated malicious domains blacklist. The are motivated by detecting and preventing drive-by-downloads, used by bot herders to infect new machine and expand their botnets \cite{978-981-15-4825-3_6}. They tested the method on offline packet capture (pcap) file containing malicious domains traffic, as well as live traffic on campus and concluded that they can detect malicious traffic in realtime \cite{ghafir2015dns}. However, while black listing is low cost and a fast solution, it is not generally effective against new and previously unknown domains.

Zhao et al, employed two phases in their detection efforts. First, preclassify URLS based in Euclidean edit distance from blacklisted domain URLs, Next, word segmentation algorithms were used to expand the URL lexical feature sets by dividing it into n-gram substrings. Weighted frequency of occurrence of n-grams whitelisted/blacklisted substrings are used to calculate a reputation score. The score was eventually used to classify the URLs before visiting the website. Realtime data was used to prove the efficacy of the proposal \cite{zhao2019malicious}.

One way of analyzing a network or dataset is by using a giant dataset of DNS queries and a blacklist, spatial co-occurrence is searched for to predict that domain names are probably malicious. s by using any of the Python modules NetworkX 4 or graph-tool. It seems the model proposed during this paper is able to search out malicious domain names that weren't within the initial blacklist, of that 28 \% had not been found by any on-line scanners before. Eventually, the results of the model are often used for additional analysis and might be included as a part of a broader detection system. However, the matter with these tools is that each one graphs are being stored in-memory and this is able to need way an excessive amount of memory for a standalone machine with these datasets \cite{zwaan2016malicious}.

Botnets "relied on Dynamic Domain Name Services (DDNS) to build Command and Control (C\&C) centers". Agyepong et al. proposed a system to detect one of the most difficult DDNS techniques, Domain Generation Algorithm (DGA), using frequency analysis of the domain names characters. They concluded that URLs generated using DGA usually have a weighted score of less than 45 \cite{agyepong2018detection}. 

Messabi et al. proposed to detect malicious sites without visiting them using 8 URL features. Some of the most common features useful in identifying malicious domains included TLD, number of dots, numerical character, total URL length and hyphens. Five weeks of real traffic traces were used to evaluate the technique using Weka. Ten-folds cross-validation tests showed J48 classifier accuracy of  77.8\%  with 80.5\% Area under the Curve and 21.4\% false positives rate \cite{messabi2018malware}.

Zhao et al. also proposed another detection techniques based on blacklisting and then classification based in a reputation score calculated from URL lexical features. They divided the URL, excluding TLD, into n-grams substrings and then computed the URL reputation based on the frequency of these substrings in the white and blacklists. They reported 94.16\% accuracy with 4.91\% false positives ratio and higher performance compared to other techniques \cite{zhao2019malicious2}.

In summary, blacklisting has proven noneffective over the time, it can however be supplementary and useful during training stages, On the other hand, studies using weighted features quantification are limited in scope and not effective against non English IDN domains. More recently attackers have benefited from the introduction of IDN domains to obfuscate their URLs and avoid detection. In this work we propose new IDN based features that would improve IDN domain detection.

\section{Proposed Approach}
This section describes the approach to DNS malware detection, the data group classification process, the features used to detect malicious domain names, and includes a detailed explanation of the experiments performed. The proposed approach to classify websites into benign and malicious, draws on the existing methods. Therefore, the algorithm combines the most potent DNS and domain name based from the literature, with novel new feature based on our datasets studies and observation. 

\begin{table} \label{T1}
	\caption{Enhanced Domain Names Statistical and Characterization Features}
	\begin{tabular}{|l|l|}  \hline	
		\textbf{Feature Set}   &   \textbf{Statistical and  }  \\   
		&   \textbf{Characterization Feature}  \\ \hline 
		&    \\ 
		&  Alphabetical Characters    \\	
		&  Number of Dots    \\			
		Basic Features	&  Number of Hyphens    \\	
		&  Digits  \\
		&    \\ \hline
		&    \\ 
		Top-Level Domains	& Unpopular TLD    \\	
		&    \\ \hline
		&    \\ 
		Character Indicator Values	& Frequency of Characters    \\	
		&    \\ \hline
		&    \\ 
		&  Unethical Words \\ 
		Tokens		&  White List  Appearance   \\	
		&    \\ \hline
		&    \\ 
		Internationalized Domain Name	& Cyrillic and Greek Alphabets   \\	
		&    \\ \hline
		&    \\ 
		Domain Age			& Domain Register Date   \\	
		&    \\ \hline
		&    \\ 
		Scanning Tool		& Reputation Rate   \\	
		&    \\ \hline								
	\end{tabular} \\
\end{table}

\begin{figure}[hb]
	\centering
	\includegraphics[width=0.9\linewidth]{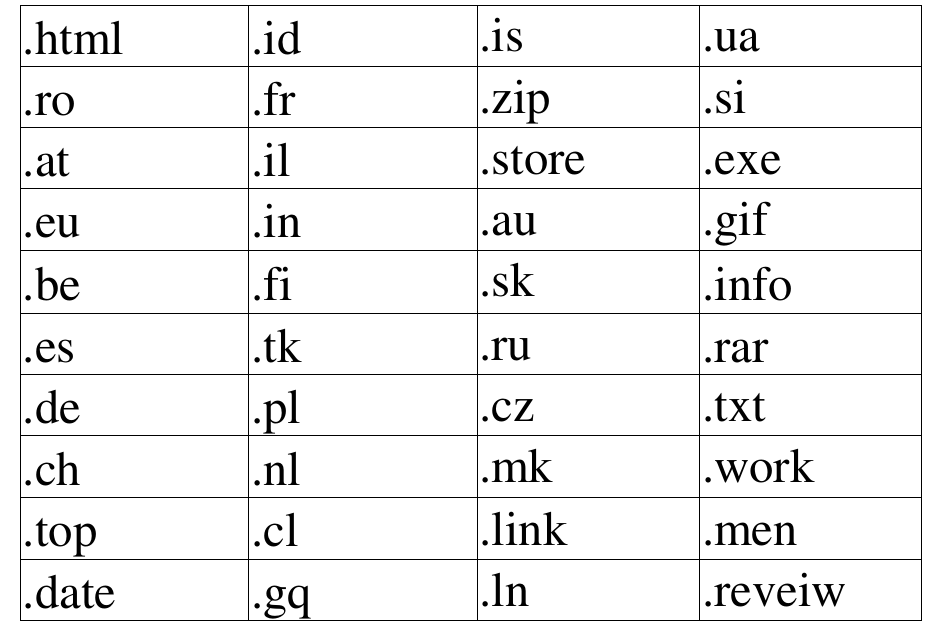}
	\caption{Suspicious Top-Level Domains \cite{wang2015breaking}}\label{TLD}
\end{figure}

\subsection{Features Selection}	

We followed the example of \cite{sharmaefficient} to select the most effective features for malicious domain detection in addition to our observations and measurements on the datasets. Table I shows the set of identified features that can be used for domain classification. The features are divided into seven categories and they are as follow: Basic features, Top-level domains-based features, character indicator values, tokens \cite{wang2015breaking}, Internationalized Domain Name, domain age \cite{domains2016}, and additional features that we introduced by using a scanning tool and gave it a rating. The following subsections introduce the basic as well as the new advanced features we use in this work.

\subsubsection{Basic Features}	
Basic features based on \cite{wang2015breaking}, and includes the number of characters, which is obtained directly as the total number of characters used, i.e. length of the domain name. The longer the length or the number of characters, can be used as an indicator for the more likelihood being a malicious website. This is due to the fact that much longer domain names are less likely to be used by legitimate websites, more likely to be available and easier to register. The other criteria is the number of dots in the domain, where dots appearing more than three times can have more potential for a malicious domain. Another basic feature is the number of hyphens, where \cite{wang2015breaking} work found that "most legitimate domains usually do not contain more than a hyphen or two". Finally, the excessive use of digits in domains can be another indicator of maliciousness.

\subsubsection{Top-Level Domain Features}	
Statistical analysis have shown that certain Top-Level Domains have more potential to be used by malicious domains than others. Figure \ref{TLD} \cite{wang2015breaking} shows a summary of the most common suspicious TLDs. This does not mean that any domain with these TLDs is classified as malicious, but it may be used as indicator of higher risk of being malicious. 

\subsubsection{Character Indicator Values'  Features}	
Another important measure, is the frequency of characters, where, malicious URLs potentially contain more repeated characters. That is the probability of maliciousness is correlated to the probability of repeated characters.

\subsubsection{Token Features}	
Domain names are also analyzed for the occurrence of certain tokens, such as unethical words, obscene words, aggressive sexual content, and illegal content \cite{wang2015breaking}. In addition, some words contained in blocked websites and those indicative of Spam emails, are also used as a good indicator for malicious domain names. Another criteria, is whether a token is listed in a whitelist database or not, and whether it belongs to a known websites or not. This is achieved by inspecting available databases, such as \cite{RiskAnalytics2019,PhishTank2019}.

\subsubsection{Internationalized Domain Name Features}	
This work presents unique features that are based on Internationalized Domain Names, where Cyrillic and Greek alphabets are used to create fake domains with characters inherited from these alphabets. This usually occur when threat actors intentionally register domains with obscure Cyrillic and Greek characters to avoid English based detectors as well as add localized legitimacy to their domains \cite{elsayed2018large}. In addition, these alphabets contain several letters that are similar to English but not the same. Therefore, they can be used to create fake malicious phishing websites that look identical to authentic ones. For instance, Table \ref{T2} shows examples of such  different Unicodes for similar characters.

\subsubsection{Domain Age Feature}	
Domain age is another important indicator, where the older the domain, the lower the risk it is to be malicious \cite{domains2016}. 

\subsubsection{Scanning Tools' Features}	
The last unique feature used in this work is a scanning tool that scans the domain and provides a rate based on reputation. The rate is used as another indicator to detect whether the domain name is malicious or not. The list of scanners used are listed in the next section.

\begin{table} \label{T2}
	\caption{ UTF-16 for Cyrillic and Greek Alphabets with Corresponding English }
	\begin{tabular}{|l|l|l|l|}  \hline	
		\textbf{Cyrillic/Greek}   		& \textbf{UTF-16}  &   \textbf{English}  & \textbf{UTF-16}   \\   \hline	 
		$A$			&  \textbackslash u0391 & $A$  & \textbackslash u0041  \\  \hline	
		$B$			&  \textbackslash u0392 & $B$  & \textbackslash u0042  \\  \hline	
		$Z$			&  \textbackslash u396	& $Z$  & \textbackslash u005a  \\  \hline	
		$\kappa$	&  \textbackslash u03ba	 & 	$k$	& \textbackslash u006b	 \\   \hline					
		$\upsilon$	&  \textbackslash u03c5	 & 	$v$	& \textbackslash u0076	 \\   \hline						
		$\omega$	&  \textbackslash u03c9	 & 	$w$	& \textbackslash u0077	 \\   \hline							
		$\iota$		&  \textbackslash u03b9	 & 	$i$	& \textbackslash u0069	 \\   \hline						
		$\nu$		&  \textbackslash u03bd	 & 	$v$	& \textbackslash u0067	 \\   \hline						
		$\chi$	    &  \textbackslash u03c7	 & 	$x$	& \textbackslash u0078	 \\   \hline						
		$\beta$		&  \textbackslash u03b2	 & 	$B$	& \textbackslash u0042	 \\   \hline						
		$\epsilon$	&  \textbackslash u03b5	 & 	$E$	& \textbackslash u0045	 \\   \hline						
		$C$	&  \textbackslash  u0421	 & 		$C$	&	\textbackslash u0043 \\   \hline							$O$	&  \textbackslash  u041E	 & 		$O$	&	\textbackslash u0043f \\   \hline						

	\end{tabular} \\
\end{table}

All of these measures, including domain name contents, statistical features, and characterization features are combined together in order to identify malicious domains. In the next subsection, we introduce the detection steps.

\subsection{Malicious domains Detection Steps}
\begin{figure*}[!hbt]
	\centering
	\includegraphics[width=0.8\linewidth]{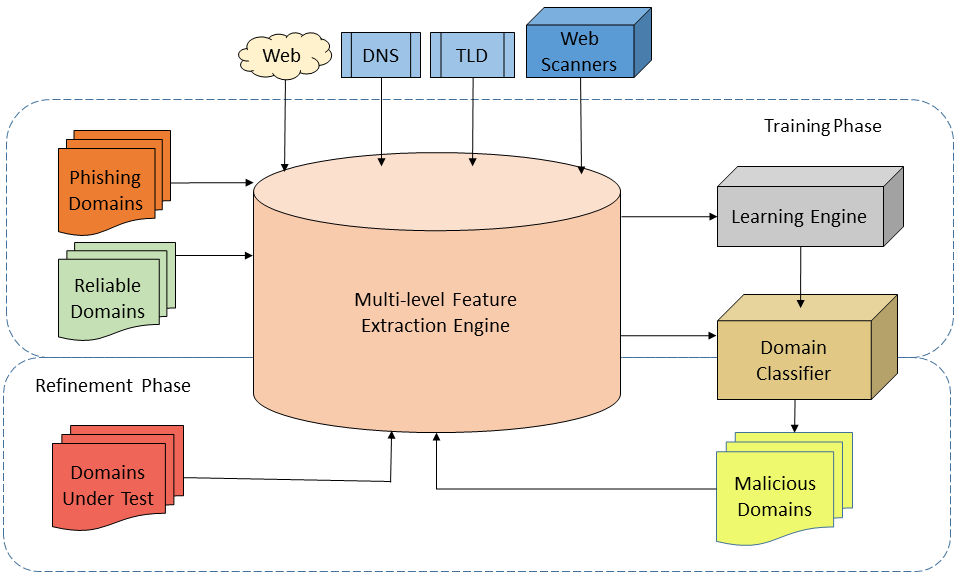}
	\caption{Malicious Domains Detection Method }\label{MLD}
\end{figure*}

Figure \ref{MLD} demonstrates the method used for features extraction and classification of the domains.  The process of extracting malicious domain names based on the selected features has been implemented using the following steps.

\subsubsection{Step 1}	
Malicious domain names were collected from different databases, including DNS-BH - Malware Domain Blocklist by RiskAnalytics \cite{RiskAnalytics2019}, and Phish Tank \cite{PhishTank2019}.

\subsubsection{Step 2}
Legitimate domains were collected from the "whitelist Alexa" databases to be compared to previously collected malicious domains.

\subsubsection{Step 3}	 
Five of the scanners for malicious domains were used in order to provide a rate out of 5.

\begin{enumerate}
	\item  	https://www.virustotal.com/en/
	\item 	https://sitecheck.sucuri.net
	\item 	https://app.webinspector.com
	\item 	https://global.sitesafety.trendmicro.com/result.ph
	\item 	https://transparencyreport.google.com/safe-browsing/search?hl=en 
\end{enumerate}

\subsubsection{Step 4}	
 Python scripting language was used to write a script that reads the features and compares them to malicious domains and extracts the specific features from each domain name. The following steps are used:

\begin{enumerate}
	\item Compare malicious domains with legitimate domains "whitelist". 
	\item Display the maliciousness rates (entered manually) for the domains.  
	\item Extracting the age of domain/months based on the WHOIS server, which is set-up by an authorized registrar in ICANN to provide up-to-date domains data that are registered within it.
	\item Calculate the length of the domains. 
	\item Calculate the number of dots in the domains.
	\item Calculate the number of hyphens in the domains. 
	\item Identify domains that contain unethical words.
	\item Extract domains containing uncommon TLDs.
	\item Identify domains containing duplicate characters. 
	\item Identify domains containing repeated numbers. 
	\item Extract domains that include Greek and Cyrillic characters similar to English characters.
\end{enumerate}

\section{Classification and Performance Evaluation}

Waikato Environment for Knowledge Analysis (WEKA), is available freely from the University of Waikato, New Zealand, under the GNU GPL. It is a collection of machine learning algorithms for data mining. Weka was used for the evaluation of the aforementioned features. The Python script placed the data in Excel that was converted into CVS and then Weka arrf file.
  
 Random-Forest Tree (RFT) classifier is a machine classifier that analyzes data and shows the learning outcomes. It provided the best classification results for our dataset, because it goes thorough process to improve the prediction quality for minority classes and in the data that the classifier is trying to read and translate it into output. 
 
 Ten-folds cross-validation was used during testing and training of the classifier. The average reported accuracy was 95.6\% with 3.41\% false positives rate. the Area under the ROC curve was 97.3\%.
  
The results are very promising and further investigation will be needed to check the effectiveness of the additional IDN and scanning tools features. One limitation was the complexity of data processing due to the IDN Cyrillic and Greek characters included.

\section{Conclusion and Future Work}

Identifying malicious websites and drive-by-downloaded using DNS and URL features, before the user downloads or executes the malicious scripts is of the utmost importance. Hackers have been actively trying to sabotage existing detection by using localized IDN names. In addition to the effective features of the previous work, we proposed IDN features using Cyrillic and Greek alphabets as well as scanning tools' features to achieve better accuracy and leaving no room for hackers to exploit English only domain name detection systems. More specifically, this  work contributed selected features from previous literature in addition to new categories of features based on Internationalized Domain Name features and reputation scores from five scanning tools. The results from the classification are very promising with accuracy of 95.6\%.

\bibliographystyle{IEEEtran}

\bibliography{main.bib}

\end{document}